# Ferroelastic Hysteresis in Thin Films of Methylammonium Lead Iodide

*Rhys M. Kennard, † Clayton J. Dahlman, † Ryan A. DeCrescent, ‡ Jon A. Schuller, ∥ Kunal Mukherjee, † Ram Seshadri, †+ Michael L. Chabinyc †\**

† Materials Department, University of California, Santa Barbara, CA 93106, United States

‡ Department of Physics, University of California, Santa Barbara, CA 93106, United States

∥ Department of Electrical and Computer Engineering, University of California, Santa Barbara, CA 93106, United States

+Department of Chemistry and Biochemistry, University of California, Santa Barbara, CA 93106, United States

*Corresponding Author: mchabinyc@engineering.ucsb.edu

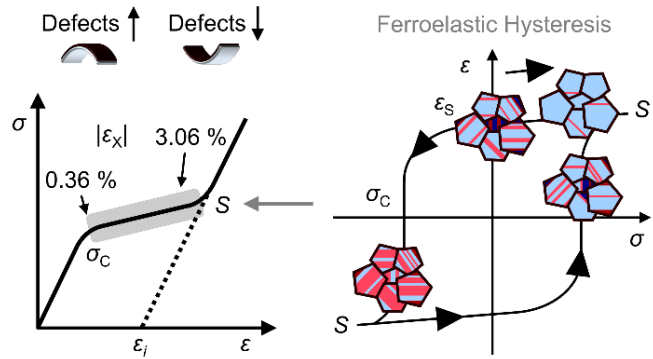

## ABSTRACT

Mechanical strain can modify the structural and electronic properties of methylammonium lead iodide $MAPbI_3$. The consequences of ferroelastic hysteresis, which involves the retention of structural memory upon cycles of deformation, in polycrystalline thin films of $MAPbI_3$ are reported. Repeatedly bent films of $MAPbI_3$ on flexible polyimide substrates were examined using Grazing Incidence Wide-Angle X-ray Scattering (GIWAXS) to quantitatively characterize the strain state, populations, and minimum sizes of twin domains. Approximate locations for the coercive stress and saturation on the ferroelastic stress-strain curve are identified, and domains from differently-strained twin sets in the films are found to interact with each other. The presence of specific twin domains is found to correlate to reports of the heterogeneity of strain states with defect content. Long-term stability testing reveals that domain walls are highly immobile over extended periods. Nucleation of new domain walls occurs for specific mechanical strains and correlates closely with degradation of the films. These results help to explain the behavior of ion migration, degradation rate, and photoluminescence in thin films under compressive and tensile strain.

## INTRODUCTION

Hybrid halide perovskites of type $APbX_3$ have emerged as materials for solar cells and are attractive for a variety of other thin film electronic devices.[1–6] The highest power conversion efficiencies in solar cells have been achieved using compositional derivatives of tetragonal methylammonium lead iodide ($MAPbI_3$), by alloying $MA^+$ with $Cs^+$ and/or $FA^+$, or by alloying $I^-$ with $Br^-$, to form $(Cs,FA,MA)Pb(Br,I)_3$.[1] Alloying of halides and introduction of large cations further provides a facile way to tune the band gap, making halide perovskites attractive for light emitting diodes or lasing applications.[3,5–10] A key advantage of halide perovskites over conventional semiconductors is the ease with which perovskites can coated via roll-to-roll processing to form electronic devices.[11] A thorough understanding of the structural consequences of repeated bending - i.e. repeated strain application - on perovskites is therefore crucial. To date, strain in $MAPbI_3$ films has been shown to affect defect concentration[12] and related properties such as degradation rate,[13,14] the activation energy of ion migration[13] and the photoluminescence lifetime.[15]

Tetragonal $MAPbI_3$ is ferroelastic which complicates how it responds to mechanical strain. Ferroelasticity involves the spontaneous formation of sub-grain domains of differing orientation, the relative proportion of which can be changed by applying external stress.[16–19] This phenomenon is common in a variety of materials, such as zirconia[20] or oxide and fluoride perovskites,[21,22] and has been increasingly investigated in halide perovskites.[23–31] At elevated temperatures many $APbX_3$ materials have higher symmetry, but as the material cools the $R_4^+$ phonon mode condenses inducing permanent out-of-phase tilting of the $BX_6$ octahedra and a spontaneous strain in the material.[32–36] The strain magnitude is lowered by transition to a lower symmetry phase, which in $MAPbI_3$ is the cubic P$m$-3$m$ to tetragonal I4/$mcm$ transition.[16,17,36] Mechanical constraints such as surrounding grains encourage twin domain formation, thus ensuring that the original macroscopic dimensions are maintained. Such domains have previously been observed in $MAPbI_3$.[24,25,29–31] Compositional derivatives of $MAPbI_3$ that are thought to be cubic on the bulk scale (e.g. $(Cs,FA,MA)Pb(Br,I)_3$) also exhibit twinning on the nanoscale.[23]

Applying stress to a ferroelastic induces ferroelastic hysteresis, which involves semi-reversibly changing the relative proportion of different domains ("domain switching") by inducing movement of the walls separating them.[18] Recent work shows that that the walls between twin domains in $MAPbI_3$ slow carrier diffusion, in single crystals without any external strain applied.[37] Despite the increased attention paid to ferroelasticity in hybrid perovskites,[23,27,28,37–42] much still remains unknown. For example, in other materials, domain walls nucleate point defects and facilitate diffusion of ionic species.[43–46] Understanding how twin walls move and are created and annihilated is therefore crucial to the successful development of flexible $MAPbI_3$-based devices.

Here, we analyze the ferroelastic hysteresis loop of $MAPbI_3$ and its impact on the stability of polycrystalline films. Cyclic strain of films revealed the approximate stresses at which ferroelastic hysteresis initiated (coercive stress) and saturated. Changes to domain sizes and proportions revealed changes in the number of domain walls and were correlated closely with enhanced degradation. This behavior was related to particular strain cycles, and can help explain differences in ion migration, degradation and photoluminescence lifetime behavior observed in literature.

## RESULTS AND DISCUSSION

*Identifying Sets of Twin Domains using GIWAXS*.

We cast MAPbI$_3$ on polyimide (Kapton) and characterized the resulting films using Grazing Incidence Wide Angle X-ray Scattering (GIWAXS) (**Figure 1a**). PEDOT:PSS was cast to planarize the Kapton surface, and was selected because it is a widely-used hole transport layer in perovskite photovoltaic devices.[47,48] MAPbI$_3$ was then spin-cast following previously-described procedures (see **SI - Methods**),[7] with an antisolvent rinse and 100°C annealing steps.[1,49] Scanning Electron Microscopy (SEM) (**Figure S1**) revealed that the MAPbI$_3$ film exhibited 200 ± 100 nm grain size and ≈ 400 nm thickness. PMMA was then cast on top of the MAPbI$_3$ to prevent degradation,[50] as subsequent experiments were largely performed in air. No X-ray scattering features of PbI$_2$ were observed (expected peak at $q \approx 0.9$ Å$^{-1}$; **Figure 1c**). The ring-like shape taken by the features in the 2D pattern (**Figure 1c**) indicates a distribution of orientations of MAPbI$_3$ crystallites, consistent with prior electron back scatter detection (EBSD) work on polycrystalline films.[51] To better understand correlations between these differing orientations and ferroelastic domains, small areas of the 2D GIWAXS patterns were integrated to form 1D patterns along two select orientations, near-in-plane (nIP) and near-out-of-plane (nOP) (**Figure 1c**). Details related to the analysis of the GIWAXS patterns are given in the **Supporting Information** (**Methods** and **Figure S2**).

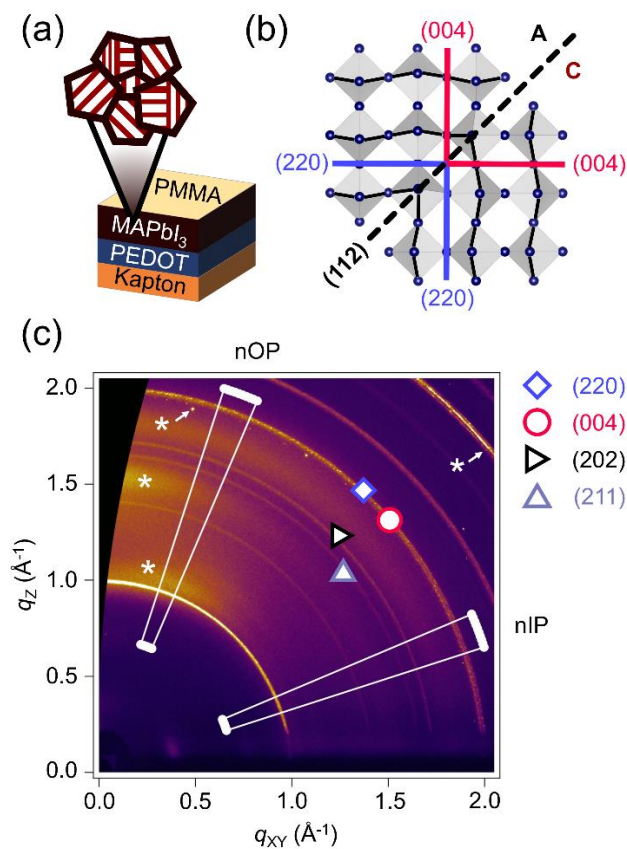

**Figure 1.** (a) Schematic of the samples as-prepared, with a Kapton substrate, a PEDOT: PSS planarization layer, MAPbI$_3$, and a PMMA encapsulation layer. The inset shows twinning in MAPbI$_3$ grains. (b)

Crystallographic structure of two twin domains of MAPbI$_3$ following reference, [24] with the (220) and (004) planes shown, as well as the (112) mirror plane (twin boundary/domain wall) separating the domains, and with MAPbI$_3$ in the I4/mcm tetragonal phase.[24,36] A and C twin types are labelled, and a sample *xyz* coordinate scheme is drawn. (c) GIWAXS pattern of an as-cast MAPbI$_3$ film, showing the areas integrated for near-out-of-plane (nOP) analysis (18°-23°) and near-in-plane (nIP) analysis (67°-72°). The (220), (004), (202) and (211) rings are labelled. The asterisks correspond to peaks observed in films with Kapton-PEDOT: PSS only (no MAPbI$_3$ or PMMA). Although appearing to overlap in this 2D pattern, the (220) and (004) peaks are distinguishable (see below).

The predominant twin domain structure and twin domain types observed by GIWAXS are shown in **Figure 1** and in **Figure 2**. At room temperature, MAPbI$_3$ is in the tetragonal I4/mcm phase,[36] with several ferroelastic domains cohabitating in the same grain (**Figure 1a**).[24,25,30] Previous Transmission Electron Microscopy (TEM) work [24] revealed that the (112) plane acts as a boundary between different domains (also called domain wall), and is also a mirror plane to these domains. Because we primarily saw ferroic *a* and *c* domain types (see below - **Figure 2**, **3**, **6**, **7**), we named the domains observed A and C. [52,53] To distinguish A vs. C, we used the scattering intensities and positions in reciprocal space of the (220) and (004) planes. The (220) nOP peak and associated (004) nIP peak revealed Domain A, while the (220) nIP peak and associated (004) nOP peak revealed Domain C (**Figure 1b**, **Figures 2-3**). However, the (220) and (004) planes have similar *d*-spacings in the 2D patterns (**Figure 1c**). For this reason, we additionally examined the evolution of the intensity of the (211) peak with respect to the (202) peak (**Figure S3**). The (211) plane is nearly parallel to the (220) plane, and the (202) plane is at equal angle from both the (220) and (004) planes, making the (202) plane a reference point. A higher ratio of intensities (211):(202) thus indicates predominant scattering from the (220) (**Figure S3**) and a lower ratio indicates predominant (004) scattering. Importantly, the analysis of the intensities of the (220) vs. (004) peaks also allowed us to extract the relative fraction of Domains A vs. C in the film, and how this proportion changes after bending.

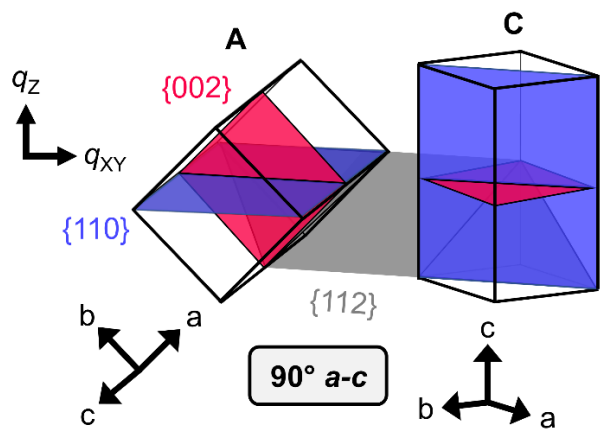

**Figure 2.** Summary schematic showing A and C domains, with preferential 90° *a-c* arrangement. [52,53]

Films that had not experienced mechanical deformation exhibited both A and C domain types (**Figure 3**). **Figure 3a** shows nOP and nIP patterns of an as-cast sample. The main nOP peak (near 3.14 Å) was assigned to (220) plane, which is consistent with a higher intensity of the (211) peak appearing nOP (**Figure 3b**, see discussion of **Figure S3**). This preference for the nOP orientation of the (220) plane is consistent with what is typically observed for MAPbI$_3$ films. [31] In the nIP pattern, the peak near 3.16 Å was considerably more intense than in the nOP direction, and the (211) peak was much weaker, indicating less scattering from the (220) peak in the nIP direction. We therefore assigned the nIP peak near 3.16 Å to be the (004) plane that corresponds to crystallites with (220) planes oriented nOP. Based on these two assignments, there is a considerable population of A-type domains in the film. Next, a clear peak was also observed in the nIP direction near 3.14 Å. This was likely a (220) peak, since the peak at this position in the nOP direction was a (220) peak. We additionally observed a peak nOP near 3.16 Å, similarly assigned to (004). These two assignments are consistent with Domain C. Thus, the typically-observed strong presence of (220) in films [31] corresponds to Domain A. The two populations identified here, Domains A and C, are consistent with the well-established 90° *a-c* configuration, hence our nomenclature for the domains (**Figure 2**).[52,53] Quantitative analysis of the relative fractions *f* of domains, and discussion of other possible configurations, required correcting the intensities for structure factor (see **Supporting Information**). We note that there were two other weaker peaks in this region, one near 3.12 Å and the other near 3.18 Å, with the latter being stronger in the nIP direction. We thus appear to have two sets of twin domains: Twin Set 1 (TS1), that is dominant, with domains A:1 and C:1, and a smaller population of Twin Set 2 (TS2).

*Correlating Strain Heterogeneity with Ferroelastic Domains*

We analyzed the strain of all peaks allowing the assignment of peaks to TS2 (**Figure 3a, 3d**). The unstrained *d*-spacing was taken from the single crystal structure of MAPbI$_3$ at 300°K.[36] To ensure consistency, we took patterns of four samples and calculated average *d*-spacings and *d*-spacing uncertainties. For TS1, the (220) peak (near 3.14 Å), hereafter called "(220):1", exhibited weak tensile strain (+ 0.08 ± 0.03 %). The (004) peak in TS1 (near 3.16 Å, "(004):1") exhibited compressive strain (− 0.37 ± 0.09 %). It is expected that the (220) and (004) planes of domains A and C for TS1 exhibit the same strain, as these domains should only differ in their orientation. For TS2, based on the improbability of having domains with strains ≥ 1.5 %, we assigned the peak near 3.12 Å to the (220) plane and the peak near 3.18 Å to (004) plane. Thus, for TS2, we have compressive strain (− 0.39 ± 0.13 %) for (220):2, and large tensile strain (+ 0.8 ± 0.4 %) for (004):2. The relative intensities of (220):2 and (004):2 roughly follow those of (220):1 and (004):1, likely indicating A and C domain types for TS2 as well, called A:2 and C:2. The strains from the (220) peaks for both twin sets are approximately a third of the magnitudes of the strains of the (004) peaks, and opposite in sign matching what is expected from Poisson's ratio of 0.33.[54] Thus, the strains identified are consistent with the film having two twin sets, each containing *a* and *c* type twin domains.

Importantly, we establish a correlation between the heterogeneity of strain in the films and the ferroelastic domains (**Figure 3d**). Sub-grain changes to strain and orientation were previously observed, but were not correlated with ferroelastic domains.[51] The *d*-spacings of the (220) peak were shown to vary by ≈ − 0.2 % within films, with the more compressed regions having higher defect densities.[12] Here, we show that variations in *d*-spacing in films correlates with different ferroelastic domains. Specifically, large compressive (220) strains originate from a different twin set. The large strains and small population of TS2 suggest that TS2 originates from unfavorable growth conditions, caused perhaps by the substrate interface, grain boundaries, rapid solvent removal, spatial constrains imposed by neighboring large grains, etc. Thus, TS2 might occur either in small grains wedged between large grains of TS1, or within grains containing primarily TS1, but near the grain boundaries (**Figure 3d**). Due to apparent interactions between TS1 and TS2 (**Figure 7**), we propose that domain walls can indeed exist between these two twin sets.

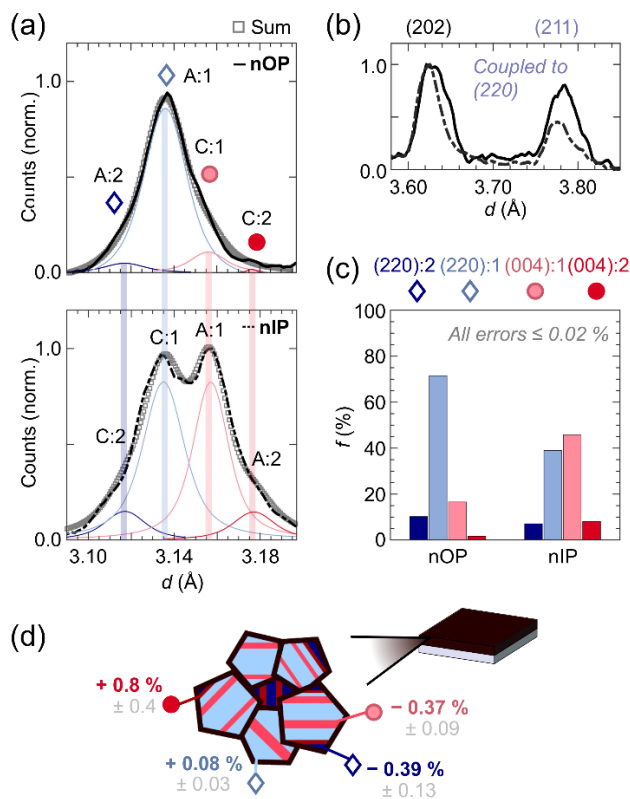

**Figure 3.** (a) GIWAXS near-out-of-plane (nOP) and near-in-plane (nIP) patterns of an as-cast MAPbI$_3$ film, with assignments to various (220) or (004) and twin domain type labelled. nOP and nIP patterns are normalized to highlight the relative contributions of different (220) and (004). Each fit is shown individually, as is the sum of the fit peaks. (b) (202)-(211) region of the nOP and nIP patterns (see **Figure S3**), normalized with respect to the (202) for (211) intensity comparison. (c) Fraction $f$ of the film exhibiting the different (220) and (004) orientations nOP and nIP for the patterns in (a). Average fractions for 4 samples are shown in **Figure S4**. (d) Summary schematic of the sub-grain twinning microstructure identified here, with two possibilities for Twin Set 2 location (separate small grains and boundaries of grains that are predominantly Twin Set 1). Strains associated with the (220) and (004) peaks are listed; and were calculated from the average and standard deviation for 4 samples.

*Volume Fractions and Minimum Sizes of A and C Domains*

Next, we calculated the relative fraction $f$ of each domain in the film (see **Supporting Information**). The fractions for the patterns shown in **Figure 3a** are shown in **Figure 3c**, and the averaged fractions for four samples are shown in **Figure S4**. We distinguish two fractions $f$, obtained respectively from the nOP and nIP patterns. The distinction is necessary, as nOP scattering detects most planes parallel to the substrate; however, nIP scattering can miss a significant number of planes perpendicular to the substrate (**Figure S5**), in crystallites that otherwise have nOP planes visible. Thus, we extract volume fraction from the nOP patterns, and use the nIP patterns to qualitatively confirm our interpretations.

From the nOP pattern (**Figure 3a**), ≈ 70 % of the film exhibits (220):1, with ≈ 20 % of the film exhibiting (004):1 and another ≈ 10 % exhibiting some combination of (220):2 and (004):2, with a greater amount of (220):2 (**Figure 3c**). This confirms the predominance of A-type domains. The results in **Figure 3** were also reproduced for 3 subsequent films, with between 80-100% of fractions *f* from the nOP patterns belonging to TS1, and the remaining 0-20 % being TS2 (**Figure S4**). Analysis of nIP patterns revealed ≈ 45 % and ≈ 40 % for the (004):1 and (220):1 respectively, in agreement with both 80-100 % of the film being TS1, and with there being some invisible peaks nIP (**Figure S5**). Some of the lack of agreement between nOP and nIP fractions could also originate from the grains being rotated in-plane with respect to each other. Another possibility is that we have some 90° *a-a* twinning, involving two A-type domains, where the domain wall belongs to the {112} family. This 90° *a-a* configuration is the structural equivalent of the 90° *a-c* configuration (**Figure 2**), but with a different orientation.[53] Prior work reported 90° *a-a* configuration for films cast directly on TEM grids,[24] rather than the predominant 90° *a-c* configuration observed here under different growth conditions. To circumvent questions of nIP plane invisibility vs. in-plane grain rotation vs. of 90° *a-c* / 90° *a-a* twinning type, we restrict quantitative analysis of fractions to the nOP patterns. Overall, the films are 80-100% TS1 and 0-20 % TS2, with a preference for 90° *a-c* twinning type (**Figure 2**, **3d**).

Finally, we calculated the minimum domain size $D_{min}$ associated with each peak using Scherrer analysis (**Supporting Information**, **Figure S6**).[55] From analysis of the breadths of the (220):1, (004);1, (220):2, and (004):2, we extracted the minimum sizes $D_{min}$ of the corresponding domains (A:1, C:1, A:2 and C:2 respectively). For A:1, C:1, A:2 and C:2, the $D_{min}$ are 38 ± 4 nm, 39 ± 9 nm, 48 ± 11 nm and 54 ± 24 nm respectively. These values represent the average and standard deviation over 4 separate samples for each peak. Large variabilities in domain size were previously observed in MAPbI$_3$,[24,25,29,30] and ≈ 40 nm is consistent with the size of smaller domains seen in films.[25,30] The grain size here is 200 ± 100 nm (**Figure S1**); so the domains are indeed sub-grain.

*Residual Stress in As-Cast MAPbI$_3$ films*

Because the impacts of thermal stress from the substrate have garnered much interest,[13,14,56] we also quantified the nIP stresses of the four domains (**Figure 4**). We multiplied the strains identified for (220):1, (004):1, (220):2 and (004):2 (C:1, A:1, C:2 and A:2, nIP) by the Young's modulus for MAPbI$_3$ (14 GPa, chosen to be mid-range among reported values) and plotted these stresses in **Figure 4**.[54,59–61] Because the modulus is nearly directionally isotropic,[61] we multiplied the strains by the same modulus for (220) and (004) values to get the nIP stresses. These stresses were 11, − 52, − 55 and 112 (± 10) MPa for (220):1, (004):1, (220):2 and (004):2 respectively. In prior work, switching from a silicon substrate with low thermal expansion coefficient or T.E.C. (0.26 × 10$^{-5}$/K) to a polycarbonate substrate with similar T.E.C. (6.5 × 10$^{-5}$/K) to those reported for MAPbI$_3$ (4-16 × 10$^{-5}$/K)[57,58] was reported to greatly reduce the average residual in-plane stress, as measured via wafer curvature.[14] However, the contributions of individual twin domains to this stress were not elucidated. Here, we have a Kapton substrate (T.E.C. 3-11 × 10$^{-5}$/K)[62,63] and PEDOT: PSS planarization layer (T.E.C. 5 × 10$^{-5}$/K),[64] with similar T.E.C. to MAPbI$_3$ (4-16

× $10^{-5}$/K), [57,58] and a PMMA capping layer that also has similar T.E.C. (5-10 × $10^{-5}$/K). [65] Because T.E.C.$_{\text{Kapton/PEDOT: PSS}}$ ≈ T.E.C.$_{\text{Polycarbonate}}$ and because the films in prior work [14] were also processed at 100°C, we compared the individual domain stresses to the reported average stress of MAPbI$_3$ on polycarbonate (12 ± 2 MPa). With the exception of the (220):1, all of the individual domain stresses exhibited much larger magnitude than the average stress of MAPbI$_3$ on polycarbonate, with some exhibiting opposite sign. Notably, most individual domain stress magnitudes were of equal or larger magnitude to the average stress reported for MAPbI$_3$ on Si (54 ± 8 MPa). These results demonstrate that the residual stresses of individual twin domains in MAPbI$_3$ can be quite large even when using a substrate of similar T.E.C..

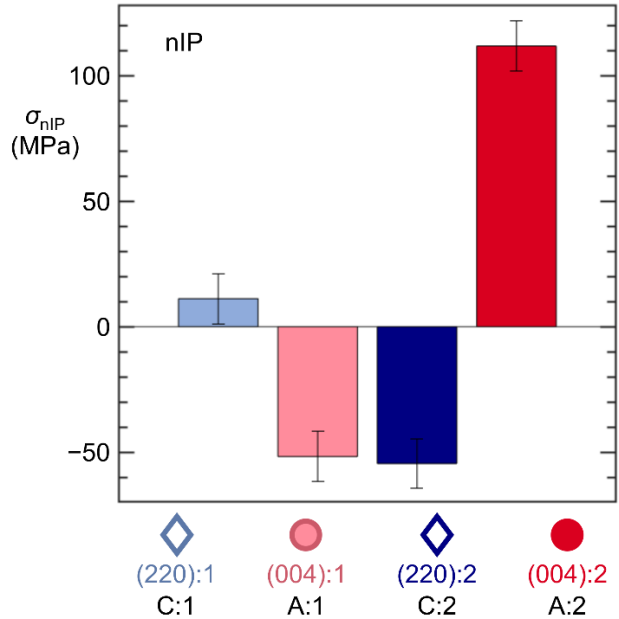

**Figure 4.** Conversion of the strains of (220):1, (004):1, (220):2 and (004):2 to stresses, with the corresponding nIP twin domains labelled. All error bars refer to the standard deviation from 4 samples.

*Expected Ferroelastic Behavior upon Bending and Experimental Design*

The ferroelastic hysteresis loop for MAPbI$_3$ has not yet been characterized in detail. Therefore, we analyzed structural changes caused by repeated bending, which is of interest for the behavior flexible devices. We will first describe expected ferroelastic behavior (**Figure 5**), then discuss the effects of repeated bending on film structure (**Figures 6,7,8**) and stability (**Figure 9**). We then relate our findings to existing literature on defect behavior in MAPbI$_3$, as domain walls have been found to nucleate vacancies and facilitate ion diffusion in other materials.[43–46]

Ferroelastic hysteresis involves non-linear, but to limited extent, reversible switching from domain A to domain C (**Figure 5a**). [18,66] This is accomplished by applying tensile strain (or stress) in $q_{XY}$, which causes the domain wall to move, provided that the applied stress has some minimum magnitude corresponding to the coercive stress $\sigma_C$. Thus, the bond distances between the Pb$^{2+}$ ions and the I$^-$ ions change *at the domain wall, on the side of domain A*, such that eventually the octahedral tilting of domain C *at the domain wall* becomes more energetically favorable. [18] The domain wall thus advances through A and converts all the octahedral tilting to that of C.

Correspondingly, the orientations of the (220) and (004) planes become that of C. If the material is under an opposite (compressive) stress in $q_{XY}$, the octahedral tilting of A becomes again more favorable, so the domain wall moves back through C and A becomes bigger. Thus, while being somewhat reversible, the process is highly non-linear (**Figure 5b**), and the domain sizes after a full ferroelastic hysteresis loop may not be the same as they were initially. Saturation $S$ (**Figure 4b**) occurs when all possible switching to either A or C has occurred. Domain switching also imparts an inelastic strain $\varepsilon_i$ on the material [20,21] that is retained after the applied strain is removed. This retention enabled us to indirectly analyze the ferroelastic hysteresis loop.

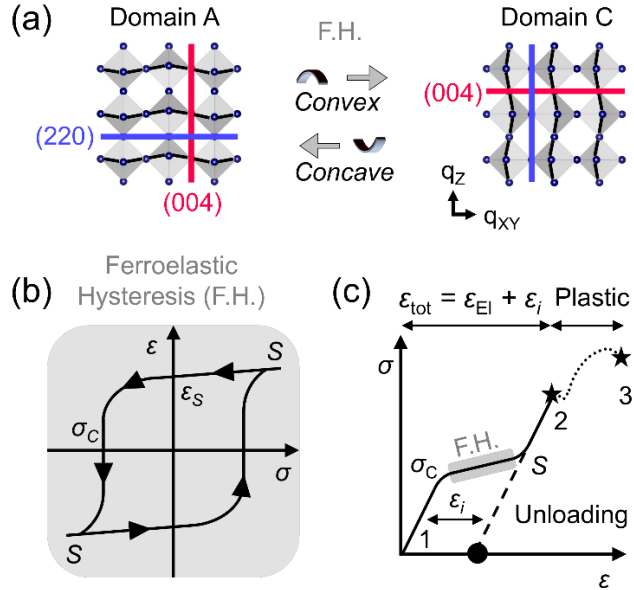

**Figure 5**. (a) Expected structural changes during ferroelastic hysteresis and plastic deformation; tetragonal crystal structure from reference.[36] *Convex* bending applies tensile in-plane strain and *concave* bending applies compressive in-plane strain. (b) Ferroelastic hysteresis loop and (c) Idealized stress-strain curve for a ferroelastic based on prior work,[20–22] where $\varepsilon_S$, $\varepsilon_i$, $\varepsilon_{tot}$, $\varepsilon_{Elastic}$ are the spontaneous, inelastic, total and elastic strains respectively, $S$ is the saturation point, and $\sigma_C$ is the coercive stress. The stars indicate fracture point locations for different materials. GIWAXS patterns were collected after unloading the applied stress (black dot on the stress-strain curve).

Ferroelastic hysteresis is one part of the stress-strain curve for ferroelastics. An idealized curve based on the literature for various oxide perovskites and zirconia is presented in **Figure 5c**.[20–22,67] Upon fabrication, a spontaneous strain $\varepsilon_S$ exists in the material. At very high applied stresses (or strains), the material is in a fully plastic regime (points 2-3). For applied stresses below the fully plastic regime, the total strain in the material is $\varepsilon_{tot} = \varepsilon_{Elastic} + \varepsilon_i$; where $\varepsilon_{Elastic}$ is elastic strain and $\varepsilon_i$ is the inelastic strain. At very low applied stresses (between point 1 and $\sigma_C$), there is an elastic-only regime, in which the domain walls do not move and bonds between atoms simply stretch or compress. At the coercive stress $\sigma_C$, ferroelastic hysteresis begins (**Figure 5b-5c**). In this regime, the types of strains acquired by the material are assigned differently from study to study; [20–22] but it is generally agreed that ferroelastic hysteresis imparts inelastic strains on the material, which for simplicity, we call the inelastic strain $\varepsilon_i$. The end of the ferroelastic hysteresis regime is marked by saturation $S$, at which point all possible domain switching has occurred. After this ferroelastic regime, a second mostly-elastic regime (until point 2) occurs, followed by the fully plastic regime and fracture. Some ferroelastics exhibit large plastic regimes, [20] others less so, [21,22] indicated by two possible locations for fracture. Experiments on free-standing $MAPbI_3$ films identified only a single mostly-linear elastic regime before fracture. [68] This indicates 1) that that the fully plastic regime is likely quite small and 2) that the slopes both quasi-elastic regimes and the ferroelastic hysteresis regime are likely quite similar. In **Figure 5c**, we have drawn them to be very different for clarity. For any stress larger than $\sigma_C$, if the applied stress is removed ("Unloading" in **Figure 5c**), the material does not go back to its initial state, due to the inelastic strain $\varepsilon_i$. Subsequent re-loading occurs much more closely to the Unloading (dashed) line in **Figure 5c** than to the original curve. In the following experiments, we focus particularly on the unloading/re-loading behavior to indirectly study the hysteresis loop.

The bending experiment is schematically shown in **Figure S7**, and referenced in **Figure 5a.** The $MAPbI_3$ stack was repeatedly bent around cylinders and then released, with two different bending configurations used. Following the naming convention used in prior works, [13–15] *convex* bending involves having the Kapton substrate touch the cylinder, with $MAPbI_3$/PMMA on the outside, and *concave* bending involves having $MAPbI_3$/PMMA touch the cylinder with Kapton on the outside (**Figure 5**, **S7**). Thus, strain $\varepsilon_X$ is applied along $q_X$, and strains $\varepsilon_Y$ and $\varepsilon_Z$ are induced in $q_Y$ and $q_Z$, following the coordinate system of **Figure S7**. Convex bending results in tensile $\varepsilon_X$ (compressive $\varepsilon_Y$ and $\varepsilon_Z$) and concave bending results in compressive $\varepsilon_X$ (tensile $\varepsilon_Y$ and $\varepsilon_Z$). Several bending diameters were selected and the applied strains $|\varepsilon_X|$ were calculated following the methodology outlined in the **Supporting Information**.[54,59–61,69–71] Using Poisson's ratio (0.33),[54] the induced strains $|\varepsilon_Z|$ in $q_Z$ were then approximated (**Table 1**). We verified via SEM that applying these strains did not cause readily observable fracture of the films (**Figure S8**). Prior to bending, the film was isotropic in-plane, as the small thermal expansion mismatch between the polymer substrate (T.E.C. 3-11 × $10^{-5}$/K, [62,63]) and $MAPbI_3$ (T.E.C. 4-16 × $10^{-5}$/K, [57,58]) induced a mild biaxial in-plane strain. We collected GIWAXS patterns immediately after the applied strain was relieved, i.e. after "Unloading" (**Figure 5**), and we call these post-bending samples "unbent".

| Diameter (mm) | $|\varepsilon_X|$ (%) | $|\varepsilon_Z|$ (%) |
|---|---|---|
| 35 | 0.34 | 0.10 |
| 31 | 0.39 | 0.12 |
| 18 | 0.67 | 0.20 |
| 10 | 1.26 | 0.41 |
| 4.1 | 3.06 | 1.01 |

**Table 1.** List of diameters employed and the corresponding applied strains $|\varepsilon_X|$ and induced strains $|\varepsilon_Z|$ to the $q_X$ and $q_Z$ directions respectively.

*Sweeping Through the Hysteresis Loop with Different Bending Diameters*

Bending with a diameter of 10 mm (**Figure 6**) induced clear signatures of ferroelastic hysteresis. **Figure 6a** shows the evolution of the nOP patterns of a film after up to 12 *convex* bending cycles around a 10 mm diameter. Following this, the film was bent *concavely* up to 12 times. After *convex* bending, the initial (220):1 intensity decreased, in favor of the (004):1 intensity increasing. Subsequent *concave* bending restored the (220):1 intensity. These observations were backed by quantitative analysis of the fractions for all peaks (**Figure 6b**) and by the decrease (*convex* bending) then increase (*concave* bending) of the (211) intensity (**Figure S9**). Thus, A:1 domains were replaced by C:1 domains upon *convex* bending, and this process was generally reversed upon subsequent *concave* bending. However, the strong retention of (220):1 intensity (**Figure 6a-b**) and the lack of clear change in minimum domain size (**Figure S10**) suggest that the domain walls did not move very far.

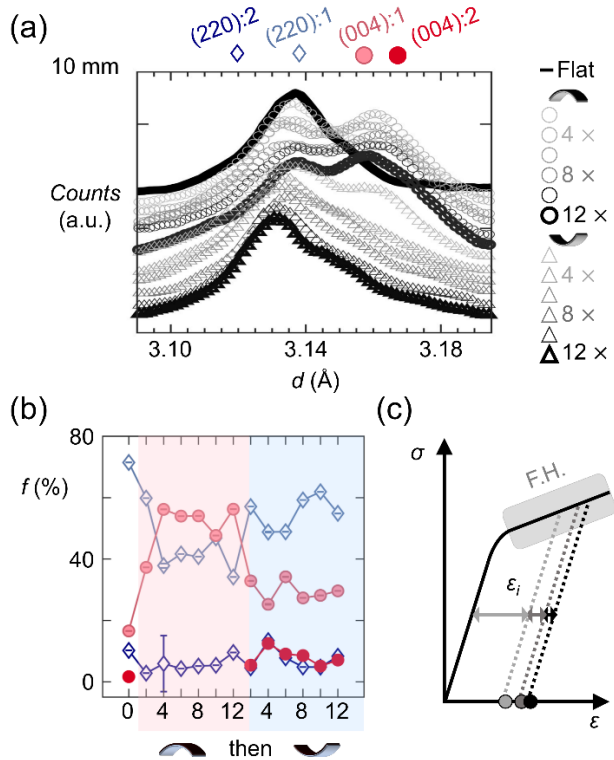

**Figure 6.** 10 mm bending experiment. (a) GIWAXS nOP patterns in the (004)-(220) region for the successive *convex* and *concave* bending around a 10 mm diameter. For the (220):1, (004):1, (220):2 and (004):2 peaks in these patterns, (b) corrected fractions *f* of the total scattered intensity in the peaks from (a). (c) Relationships between the observed domain switching and the stress-strain curve.

The changes observed can be correlated with the stress-strain curve, shown in **Figure 6c**. In cycles 4-12 for both *convex* and subsequent *concave* bending, the fractions plateaued at consistent values (**Figure 6b**), indicating that after cycle 4, the amount of A vs. C was stable. This suggests that during the first *convex* bending, $MAPbI_3$ followed the initial stress-strain curve, and the A:C domain ratio changed. When the applied stress was removed, the curve followed the unloading line (dashed). Subsequent re-loading and unloading occurred in a similar location to the first unloading line, leading to the plateau in the fraction of each domain. During subsequent *concave* bending, the domain walls moved closer to their original positions, and the unloading/reloading lines move closer to the original curve.

We next applied larger strains to MAPbI$_3$ films. The bending experiment was repeated with a diameter of 4.1 mm ($|\varepsilon_X|$ = 3.06 %; $|\varepsilon_Z|$ = 1.01 %; **Figure 7**). Separately, we investigated the effects of *concave* bending only, without prior *convex* bending, which are distinguished in **Figure 7a-b** by the gray shadowing. The changes to the nOP-patterns were similar to that of the 10 mm sample and are discussed in the **Supporting Information (Figure S11)**. Here, we focus on key differences with respect to the 10 mm experiment as the applied strain is increased. Some sample-to-sample variability was observed, as shown by the error bars in **Figure S4** (fractions for 4 as-cast samples) and by the slight variations in fractions and minimum domain sizes for the as-cast samples (cycle 0 in **Figure 6** and **Figure 7**). However, the differences in trends between the 4.1 mm and 10 mm bending are striking enough to be attributed to the larger applied strain (**Table 1**).

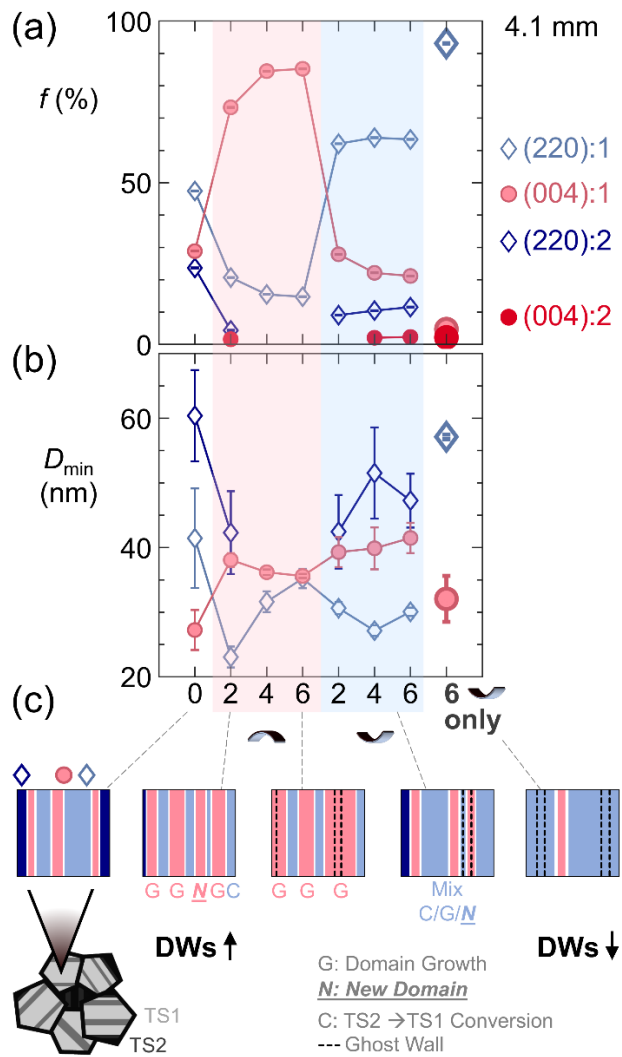

**Figure 7.** 4.1 mm bending experiment. (a) Corrected fractions of total scattered intensity $f$ (uncertainties < 0.1 %) in the (220)-(004) region and (b) changes in minimum domain size $D_{min}$ for the (220):1, (004):1,

(220):2 and (004):2 peaks of nOP patterns (see **Figure S11**). (c) Summary schematics of the processes occurring during the 4.1 mm diameter bending experiments, with the evolution in the number of domain walls (DWs) listed. TS stands for Twin Set. For clarity, only the prior locations of walls that disappeared are shown as ghosts, rather than the prior locations of *both* walls that moved and walls that disappeared.

Saturation was nearly attained upon bending around a 4.1 mm diameter (**Figure 7a**). As-cast, the film exhibited the typical preference for A:1 ((220):1 nOP), with some C:1 ((004):1 nOP) and A:2 ((220):2 nOP). *Convex* bending caused a large increase in C:1 population to ≈ 85 % of the total nOP fraction, accompanied by a large decrease in the A:1 fraction and complete disappearance of A:2. No fracturing was observed (**Figure S8**). This process was to some extent reversed with concave bending, with A:1 ((220):1) occupying ≈ 65 % of the total nOP population, which is still higher than the as-cast A:1 population (≈ 50 %). A:2 reappeared, although at lower fraction than for the as-cast sample. Strikingly, concave-only bending induced near-complete (95 %) A:1 population nOP, indicating near-complete saturation. These trends in the fractions suggest near-complete cycling through the hysteresis loop, and certainly to greater extents than for the 10 mm diameter.

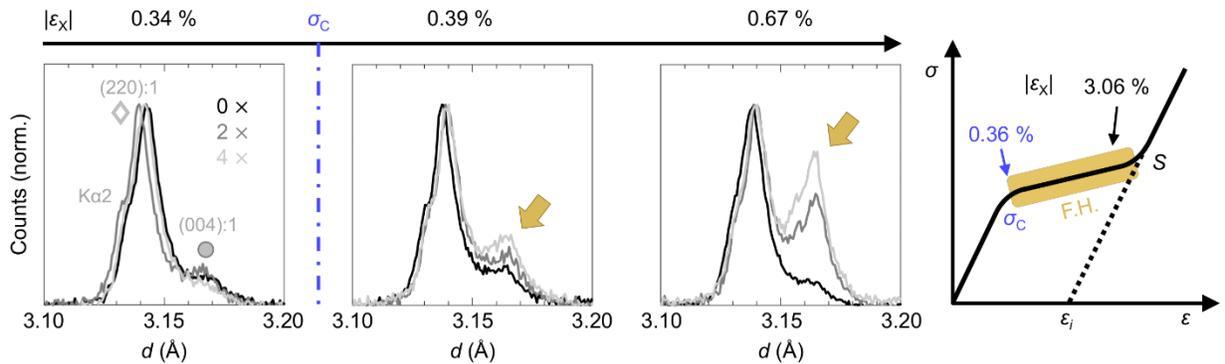

**Figure 8.** Determination of the coercive strain, from which the coercive stress σC can be calculated. Powder XRD data for the 35 mm, 31 mm and 18 mm bending diameters, corresponding to $|\varepsilon_X|$ of ≈ 0.34 %, 0.39 % and 0.67 % respectively. The location of the coercive strain on the stress-strain curve is shown on the right. The yellow arrows show growth of the (004):1, indicative of ferroelastic hysteresis (F.H.).

Finally, we repeated the bending experiment with much larger diameters, to find the coercive stress (**Figure 8**). Because (004):1 grows into a peak that is very distinct from (220):1 (**Figure 6, 7**) we were able to use powder X-Ray diffraction (PXRD), rather than GIWAXS. Due impurities in the X-Ray wavelength (Cu-Kα2, tungsten and Cu-Kβ) resulting in extra peaks, and due to the low counts on the powder diffractometer, TS2 could not be resolved as easily; however, the onset of hysteresis was clearly identified using the TS1 peaks. The PXRD patterns of films bent convexly up to 4 × around diameters of 35 mm, 31 mm and 18 mm are shown in **Figure 8**, with applied strains $|\varepsilon_X|$ of ≈ 0.34 %, 0.39 % and 0.67 % respectively. The patterns shifted slightly and inconsistently in *d*-spacing after bending. This is likely caused by the lack of height alignment in PXRD, which is done in GIWAXS and necessary for detailed strain analysis of flexible films. Notably, the (004):1 did not grow when $|\varepsilon_X|$ ≈ 0.34 %, grew noticeably when $|\varepsilon_X|$ ≈ 0.39 %, and grew significantly when $|\varepsilon_X|$ ≈ 0.67 %. A control experiment was performed to confirm that these changes were not induced by the X-Ray beam (**Figure S12**). Thus, the coercive strain is $|\varepsilon_X|$ ≈ 0.36

% Because the actual applied strain is in qX, we take |$\varepsilon_X$| to calculate the coercive stress. Using a modulus of 14 GPa, [54,59–61] the coercive stress for MAPbI3 is thus $\sigma_C \approx 50$ MPa.

### *Impacts for Controlling Twin Population during Film Growth*.

If an external stress source, such as thermal stress from the substrate or roll-to-roll printing setup, causes application of more than the coercive stress, |50| MPa in $\varepsilon_X$, it will modify the twin domain population. The stress $\sigma_T$ imparted by thermal mismatch can be calculated using $\sigma_T = \frac{E_P}{1-v_P}(\alpha_S - \alpha_P)\Delta T$, [14] where $E_P$ is the Young's modulus, $v_P$ is the Poisson's ratio, $\alpha_S$ and $\alpha_P$ are the T.E.C.s of the substrate and the perovskite and $\Delta T$ is the temperature gradient while cooling. Using the values discussed in the Residual Stress section, ($E_P$ = 14 GPa, $v_P$ = 0.33, $\alpha_P \approx 5 \times 10^{-5}$/K and $\Delta T$ = 100°C − 20°C = − 80°C), a low T.E.C. substrate such as silicon ($\alpha_S = 0.26 \times 10^{-5}$/K) will induce ≈ 50 MPa on MAPbI3 for a low annealing temperature of 70°C. A glass substrate with mid-range T.E.C. ($\alpha_S = 1 \times 10^{-5}$/K) will induce the coercive stress upon cooling from 80°C and finally, a substrate with a higher T.E.C. such as Kapton ($\alpha_S = 3 \times 10^{-5}$/K, chosen among the lower reporter T.E.Cs for Kapton) [62,63] will induce the coercive stress upon cooling from 140°C. These numbers will vary based on the $E_P$ and T.E.C. selected for the calculation; however, it is clear that ferroelastic behavior should be quite different with substrates of differing T.E.C. Thus, it does seem that thermal mismatch can be used to tailor the ferroelastic domain population. However, this does not exclude interface interactions or other solution-casting phenomena [72] which may also greatly direct twin population. Further work is needed to decouple these effects.

### *Interaction between TS1 and TS2 (4.1 mm)*.

The evolution in fractions *f* for TS2 indicated interaction between TS1 and TS2 (**Figure 7**), and therefore, physical contact between them. The fraction of A:2 (tracked via (220):2) dropped as the fraction of C:1 increased between *convex* cycles 0-2. Moreover, both A:2 and C:2 were completely gone after 4 *convex* cycles. If domain walls moved between A:2 and C:2, we would expect a rise in the amount of C:2; and we instead saw a rise in the amount of C:1. It is possible that the large strain applied inelastically compressed the tensile strained (004):2, making it appear as (004):1. In support of this, there was more A:1 after subsequent concave bending than there was initially, so some TS2 may have been converted to TS1. However, subsequent concave bending also partially restored A:2 at the expense of C:1. This suggests that there is a mobile domain wall between C:1 and A:2, and thus, that TS1 and TS2 can be in the same grain. TS2, which exhibits larger strains than TS1, is likely to appear in areas in which local strain is greater, such as near grain boundaries or at the substrate interface (**Figure 3d** and **7c**). Since the film is densely packed with few pinholes (**Figure S1**) TS2 may also help offset strain gradients within the film.

### *Changing the Number of Domain Walls during Bending (4.1 mm)*.

We examined changes to the minimum domain size $D_{min}$ for the 4.1 mm bending experiment (**Figure 7**) in more detail, to see if the walls only moved, or if walls were created/annihilated as well. We focus on TS1, as the peaks in TS2 were quite weak. *Convex* bending induces more C, and *concave* bending induces more A (**Figures 5, 6** and **7**), due to wall movement. Thus, if we only have wall movement, C domains should become bigger after *convex* bending and A domains should become bigger after *concave* bending. However, if new C domains nucleate during *convex* bending, the size of C should decrease. These trends should also apply to the *minimum* size $D_{min}$, estimated using Scherrer broadening analysis (see above and the **Supporting Information**). Rapid growth of nucleated domains might mean that $D_{min}$ of C decreases only slightly, as opposed to sharply decreasing. With respect to the minimum size specifically, if many C domains are annihilated during *concave* bending, the $D_{min}$ of A domains will likely increase significantly. Such domain nucleation/annihilation is necessarily accompanied by an increase/decrease in the number of domain walls. In MAPbI$_3$ under applied stress, both domain nucleation and annihilation have been observed, so we know that large changes to the number of domain walls is possible. [30]

Changes in $D_{min}$ revealed both wall movement and creation/annihilation (**Figure 7b-c**). $D_{min}$ of C:1 increased after 2 cycles of *convex* bending, and $D_{min}$ of A:1 dropped. Thus, the walls moved through A:1s to increase the sizes of C:1s. This is reflected in the large increase in fraction of C:1 (**Figure 7a**). The minimum size of C:1 then decreased (cycles 4-6) while fraction of C:1 continued to increase, suggesting C:1 nucleation upon further *convex* bending, and thus, more walls. Curiously, the $D_{min}$ of A:1 concurrently increased. This suggests annihilation of the smallest A:1 domains as walls move from C:1s. The mixture of wall movement with wall creation/annihilation during *convex* bending/unbending is consistent with previous microscopy observations made on MAPbI$_3$ single crystals.[30] This process of wall movement/creation/annihilation appeared to be reversed with subsequent *concave* bending: $D_{min}$ of A:1 decreased, suggesting A:1 nucleation. $D_{min}$ of C:1 increased slightly but stayed within measurement uncertainty, suggesting slight C:1 annihilation. In the case of *concave* bending only (no prior *convex* bending), the minimum domain size of A:1 was 57.1 ± 0.4 nm, which is substantially larger than the range of as-cast domain sizes for A:1 (38 ± 4 nm, 4 samples). This result combined with the 95 % proportion of A:1 (**Figure 7a**) suggests the microstructure shown in **Figure 7c**, in which large A:1s are separated by very few C:1s, with very few domain walls total in the grain. Such a microstructure would result from progression of walls from A:1s through most C:1s such that many walls are annihilated, and large A:1s remain. Again, this behavior is consistent with observations of wall annihilation under *concave* bending in single crystals. [30] Thus, *convex* bending resulted in both wall creation and annihilation, while *concave* bending only seemed to favor annihilation.

*Domain Walls and Long-Term Stability*.

We also examined the long-term stability of films with no bending, and either 6 *convex* cycles only or 6 *concave* cycles only around the 4.1 mm diameter (**Figure 9**). We stored these films for 7 months in ambient and in the dark, and then examined powder XRD patterns. Specifically, we looked for the appearance of the degradation product PbI$_2$ at $d \approx 7$ Å (**Figure 9a**) and for major changes in the proportions of A:1 vs. C:1 (**Figure 9b**). Due to low signal on the powder diffractometer, TS2 was not easily resolvable. Prior *convex* bending only correlated with a large amount of PbI$_2$, while no bending and *concave* bending only showed no PbI$_2$ (**Figure 9a**). In addition, the patterns in the (220)/(004) region (**Figure 9b**) were quite similar to those 7 months prior (**Figure S11a**), suggesting that the proportions of A:1 vs. C:1 changed negligibly over 7 months (**Figure 9b**). Thus, without external stimulus, the domain walls appear to be fairly immobile.

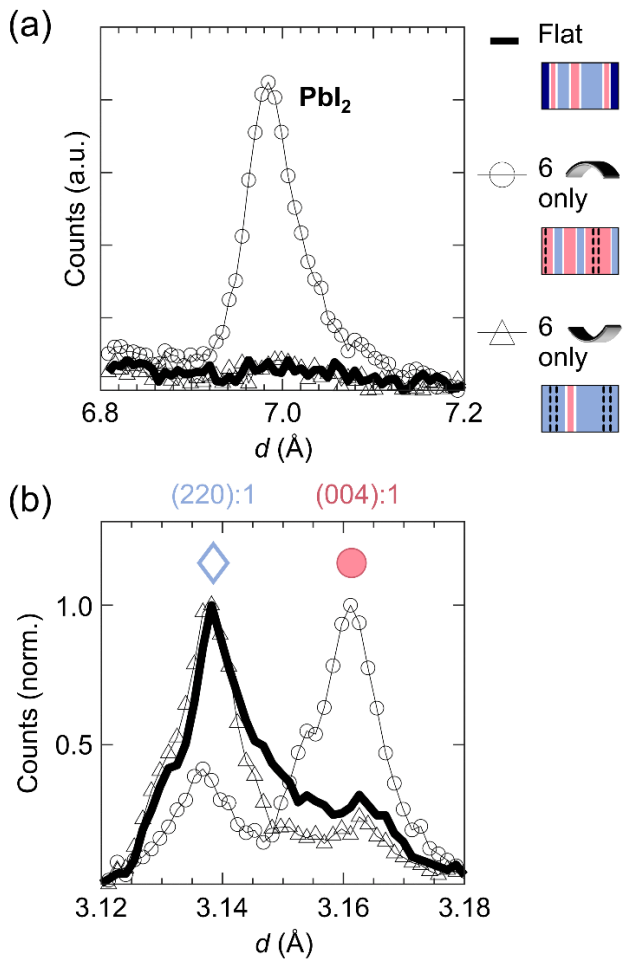

**Figure 9.** Powder XRD patterns of non-bent films (flat) and films previously bent *convexly* and *concavely* around a 4.1 mm diameter, after 7 months of storage in air, showing (a) the region in which PbI$_2$ typically appears and (b) the (220)/(004) region. Schematics are shown of twin domain layout prior to storage (see **Figure 7c**).

The difference in degradation between *convex* bending vs. no bending/*concave* bending (**Figure 9**) suggests that the walls nucleated during *convex* bending encourage degradation. As discussed above (**Figure 7**), many domain walls seemed to nucleate during *convex* bending, with some walls also being annihilated. In other systems, walls are known to harbor a higher concentration of vacancies and to facilitate diffusion of vacancies/ionic species.[43–46] Domain walls can also nucleate point defects,[73] so the walls nucleated during *convex* bending may have also nucleated more point defects. However, we found that walls were also annihilated upon repeated *convex* bending. When they move, domain walls can leave behind some of the vacancies they contain, creating a "ghost line" of vacancies in the old wall location, also sometimes called "ghost wall".[43] This can create a "memory" of the wall,[74,75] which has been observed several times in MAPbI$_3$.[24,29] This memory could mean that wall annihilation does not fully compensate for the point defects created during wall nucleation. Consistent with this interpretation, both non-bent and *concavely*-bent films showed no PbI$_2$ formation, suggesting a lack of point defect creation. Thus, the domain wall nucleation during *convex* bending seems to explain the faster degradation rate.

*Discussion and Impact for Device Properties*.

Our findings that *convex* bending enhances wall nucleation and that *concave* bending encourages annihilation help explain prior defect-related results in MAPbI$_3$.[13–15] Prior studies found that films maintained under *convex* bending exhibited lower activation energy for vacancy-mediated ion migration[13] and more PbI$_2$ formation (i.e. degradation),[13,14] than flat films, suggesting higher defect content. Correspondingly, the photoluminescence lifetime of MAPbI$_3$ films decreased during *convex* bending in a different study,[15] suggesting more trap-mediated recombination. The opposite trends were observed for *in-situ concave* bending, suggesting fewer point defects/ domain walls. These results can be explained by our findings that domain walls are created during *convex* bending and annihilated during *concave* bending. Our results also indicate that repeated bending in one configuration (*convex* or *concave*) will likely not change the above-described behaviors much after the first few bending cycles (**Figure 6** and **Figure 7**). In addition, domain walls formed will not move without external stimulus. Subsequent bending in the opposite configuration (i.e. *concave* or *convex*) will change these defect behaviors to limited extents. To avoid such defect-related changes, the films should not be bent past the coercive stress (≈ 50 MPa), which is likely most relevant in the context of roll-to-roll processing MAPbI$_3$.

Changing the number of twin walls in films upon repeated bending may greatly impact carrier diffusion. Our findings reveal twin wall movement, creation and annihilation, and strongly suggest the presence of "ghost walls" [24,29,43,74,75] in the sample after bending, which is a memory of the wall formed by the vacancies the departed wall has left behind. Deep traps seem to nucleate along with twin walls *in films*, as evidenced by decreased photoluminescence lifetime upon *convex* bending (i.e. enhanced trap-mediated recombination).[15] In apparent contrast, recent work on single crystals suggests that twin walls contain shallow traps that slow carrier diffusion, but do not induce trap-mediated recombination.[37–39] It is possible that single crystals contain fewer deep traps than films, and thus that twin walls impact carrier diffusion very differently in films vs. single crystals. It is also not clear to what extent twin walls carry their traps with them when they move or annihilate. "Ghost walls" may therefore greatly impact carrier transport in both films and single crystals. The impacts of repeated bending on shallow trapping and carrier diffusion requires further investigation, particularly in films.

We note that the detailed ferroelastic loop will also depend on factors such as the grain size. In oxide perovskites, grain size is known to affect domain/domain wall formation.[76] Here, we used films with grains of 200 ± 100 nm, but films used in devices can have grain sizes of up to several microns,[77] which may affect the types of domains formed and their strains. Bringing the perovskite closer to the cubic phase via either temperature or composition should lower the spontaneous strain, and thus reduce the size of the hysteresis loop.[18,78] In addition, dislocations at interfaces are known to pin domain walls, thus preventing domain wall movement.[79,80] Complete understanding of defect behavior in MAPbI$_3$-based devices will therefore likely require investigations of how grain size, temperature, pinning, substrate interface interactions and other parameters affect the ferroelastic hysteresis loop of MAPbI$_3$.

## CONCLUSIONS

In conclusion, we have examined the ferroelastic twinning behavior in polycrystalline thin films of $MAPbI_3$ in detail. Strain inhomogeneity in the films originated from specific twin domains. We then characterized the ferroelastic hysteresis loop of $MAPbI_3$, and identified approximate applied strain values for the onset of the loop (coercive stress), partial cycling through the loop, and saturation. Nucleation of domain walls during *convex* bending correlated with enhanced degradation, and this was not offset by simultaneous annihilation of other domain walls, suggesting the presence of "ghost" walls. We also identified the temperatures at which substrates with various thermal expansion coefficients could influence the twin population by imparting stress during thermal processes. These results help to understand the structural processes related to defects in polycrystalline films of $MAPbI_3$. We anticipate that the method presented above for characterizing the hysteresis loop will aid future studies of how the loop changes under different device fabrication and operation conditions.

## ASSOCIATED CONTENT

**Supporting Information**. Experimental methods, additional characterizations including Scanning Electron Microscopy (SEM), reproduction of certain experiments for 4 samples, further GIWAXS and XRD characterizations and analyses, as well as derivations and explanation of corrections applied.

## AUTHOR INFORMATION

### Corresponding Author

*Michael L. Chabinyc: mchabinyc@engineering.ucsb.edu*Michael L. Chabinyc: mchabinyc@engineering.ucsb.edu

### Author Contributions

The manuscript was written through contributions of all authors. All authors have given approval to the final version of the manuscript. The authors declare no competing financial interest.

### ACKNOWLEDGMENT

Growth and structural characterization was supported by the U.S. Department of Energy, Office of Science, Basic Energy Sciences, under Award Number DE-SC-0012541. Support of optical characterization was provided by the U.S. Department of Energy (DOE), Office of Science, Basic Energy Sciences (BES), under Award Number DE-SC0019273. Use of the Stanford Synchrotron Radiation Lightsource, SLAC National Accelerator Laboratory, is supported by the U.S. Department of Energy, Office of Science, Office of Basic Energy Sciences under Contract No. DE-AC02-76SF00515. The research reported here also made use the shared facilities of the UCSB MRSEC (National Science Foundation DMR 1720256), a member of the Materials Research Facilities Network (www.mrfn.org). R.M.K. gratefully acknowledges the National Defense Science and Engineering Graduate fellowship for financial support. The authors would like to thank Prof. Anton Van der Ven for advice on ferroelasticity.

**ABBREVIATIONS**

GIWAXS, Grazing Incidence Wide-Angle X-Ray Scattering; $f$, fraction (by volume) of the film with a particular orientation; $D_{min}$, minimum domain size; nIP, near-in-plane; nOP, near-out-of-plane.

# Ferroelastic Hysteresis in Thin Films of Methylammonium Lead Iodide

# Supporting Information

*Rhys M. Kennard, [†] Clayton J. Dahlman, [†] Ryan A. DeCrescent, [‡] Jon A. Schuller, [ǁ] Kunal Mukherjee, [†] Ram Seshadri, [†+] Michael L. Chabinyc [†]\**

† Materials Department, University of California, Santa Barbara, CA 93106, United States

‡ Department of Physics, University of California, Santa Barbara, CA 93106, United States

ǁ Department of Electrical and Computer Engineering, University of California, Santa Barbara, CA 93106, United States

+Department of Chemistry and Biochemistry, University of California, Santa Barbara, CA 93106, United States

*Corresponding Author: mchabinyc@engineering.ucsb.edu

# Methods

*Materials*

Lead (II) iodide was purchased from Sigma Aldrich (PbI$_2$ – 99.999% purity, trace metal basis) Methylammonium iodide (CH$_3$NH$_3$I, ≥ 99% purity) was purchased from Dyseol. N-N-dimethylformamide (DMF, 99.8%, anhydrous), dimethyl sulfoxide (DMSO, ≥ 99.9%, anhydrous), Chlorobenzene (99.8%, anhydrous) were purchased from Sigma Aldrich and kept in a nitrogen glove box.

*Spin-casting of Kapton-PEDOT: PSS-MAPbI$_3$-PMMA*

Kapton sheets were cleaned via ultrasonication in isopropyl alcohol for 10 min. To stabilize the Kapton against further changes during heating, the cleaned sheets were carefully put between two aluminum plates, and the stack was heated at ≈ 350°C for 2h on a hot plate in an N$_2$-filled glove box. Following this, the heat was turned off and the stack was allowed to cool naturally to room temperature. The pre-treated Kapton sheets were then exposed to an oxygen plasma at ~ 300 mTorr for 10 min, with air as the oxygen source. To keep the Kapton flat during spin-casting, the Kapton sheets were gently placed onto a glass slide covered with a thin PDMS sheet (for information on making the PDMS sheet, see reference [1]). For annealing, the Kapton sheets were then removed from the glass-PDMS support. Care was taken to not bend the Kapton sheets during any of these steps. PEDOT: PSS (≈ 300 µL) was spin-cast in air onto the treated Kapton substrates at 2000 rpm for 10 seconds. The Kapton-PEDOT: PSS stack was annealed at 130°C for 4 min. To ensure film smoothness, 1) Kapton substrates of at least 2 cm × 2 cm size were required and 2) the PEDOT: PSS spin-casting and annealing procedures were repeated a second time, with the second annealing being 6 min. The samples were then transferred to a nitrogen-filled globe box for MAPbI$_3$ and PMMA spin-casting. The precursor solution for MAPbI$_3$ was fabricated in a nitrogen-filled glove box. PbI$_2$ and CH$_3$NH$_3$I were dissolved in 1 mL DMF and 96 µL DMSO to make a 1M solution, and the mixture was stirred overnight under mild heating (60°C). The solution was then spin-cast onto the Kapton-PEDOT: PSS at 1000 rpm for 10 s then 4000 rpm for 30 s. When 8 s passed after the spin turned 4000 rpm, 0.2 mL of anhydrous chlorobenzene was dropped on the substrate. The films were then immediately annealed at 100 °C for 10 min on a hot plate, again in a nitrogen-filled glove box. All temperatures were verified with a thermocouple. PMMA was then cast onto the Kapton-PEDOT: PSS-MAPbI$_3$ as a capping layer. 60 µL of a 25 mg/mL solution of PMMA in toluene was spin-cast onto the stack at 2000 rpm for 30 seconds, and no further annealing treatment was performed.

*GIWAXS characterization*

Grazing Incidence Wide-Angle X-Ray Scattering (GIWAXS) experiments were performed on beamline 11-3 (12.7 keV, wiggler side-station) at the Stanford Synchrotron Radiation Lightsource (SSRL). The source-to-detector (two-dimensional Rayonix MX225 CCD) distances were calibrated using lanthanum hexaboride (LaB6). All raw images were geometrically corrected using Nika. Sections (cakeslices) of the 2D GIWAXS patterns at specific angles were selected and integrated to obtain 1D patterns. GIWAXS analysis was performed primarily on two cakeslices: the first near-out-of-plane ("nOP", 18°-23° - **Figure 1c**) and the second near-in-plane ("nIP", 67°-72°). Both cakeslices were chosen to be 18° from 0° and 90°. Due to sample roughness and to avoid double diffraction issues, a large incidence angle (2°) was chosen, resulting in the near-0° angles being cut off. Nevertheless, the parameters chosen enabled clear determination of near-in-plane vs. near-out-of-plane twin orientations and of the strain magnitudes of the planes in each twin. All patterns were converted to *d* from *q*, and all peaks were fit to Pseudo-Voigt patterns using Igor, with Gaussian and Lorentzian contributions kept constant (Igor Multipeak "shape" factor of 1) across all peaks and samples. The *d*-spacings of the peaks in **Figure 3a** were then compared with measured *d*-spacings of $MAPbI_3$ single crystals at 300°K [2] to assign (220) vs. (004) nature and to calculate strain magnitude (see also discussion of **Figure S3**).

*Other characterizations*

Scanning Electron Microscopy was performed using an FEI Nova Nano 650 FEG SEM operating at 10-20 keV accelerating voltage with beam currents of 0.40-0.80 nA. For SEM measurements, the samples were sputter-coated with gold to prevent charging. No PMMA was cast on samples used for top-down measurements, to get accurate grain size measurements.

Powder X-Ray Diffraction patterns were obtained using a Panalytical Empyrean powder diffractometer in reflection mode with a Cu-Kα source, operating with an accelerating voltage of 45 kV and beam current of 40 mA.

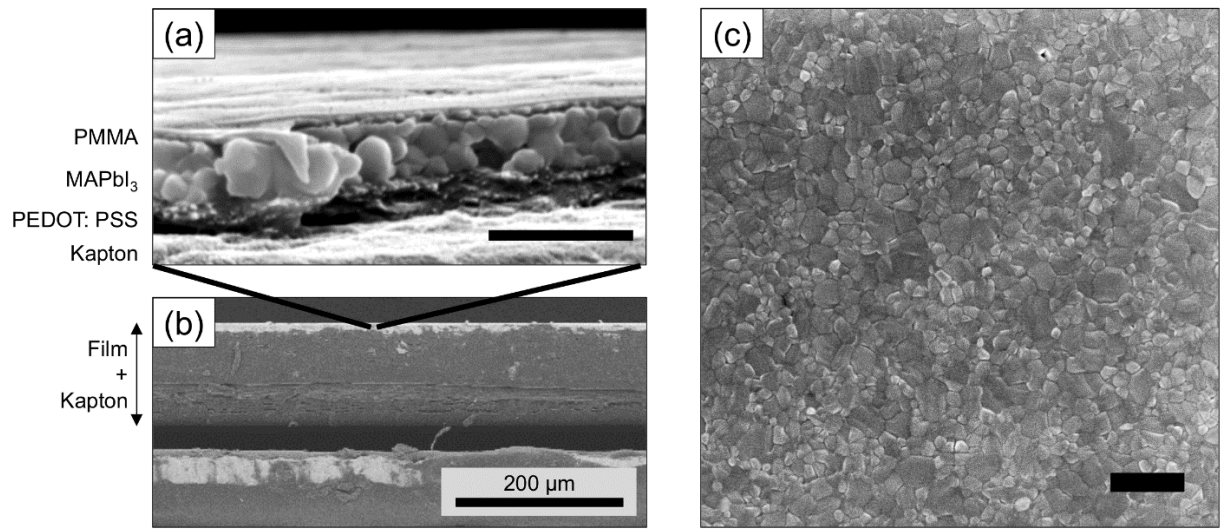

**Figure S1.** SEM images of an as-cast film, showing (a) a cross-section of an as-cast film, (b) a lower-magnification cross-section showing the Kapton thickness and (c) a top-view image of an MAPbI$_3$ film on top of which no PMMA was cast. 19 keV accelerating voltage was used for (a) and (c) and 10 keV was used for (b). Scale bars in (a) and (c) are 1 µm.

The MAPbI$_3$ film measured here was 1-4 grains thick. Because the films were cast on plastic substrates, they did not break in half cleanly, so parallax issues were encountered when trying to determine thickness. The exact thicknesses were difficult to measure, but were approximately 400 nm for MAPbI$_3$, 200 nm for PEDOT: PSS and 50 nm for PMMA. To properly measure grain size without parallax issues, we measured a film from the top (**FigureS1c**) that did not have PMMA. The average grain size was estimated from measurement of 100 grain widths to be 200 ± 100 nm.

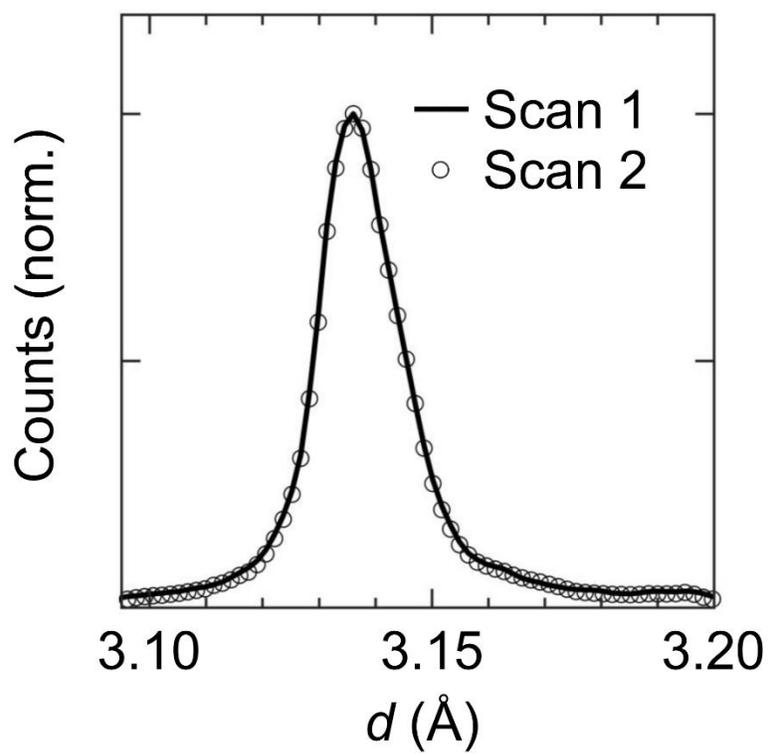

**Figure S2.** nOP patterns of films bent *concavely* around a cylinder with 4.1 mm diameter. Scan 1 and Scan 2 indicate two consecutive scans of the same sample, showing lack of change and thus good stability under the beam.

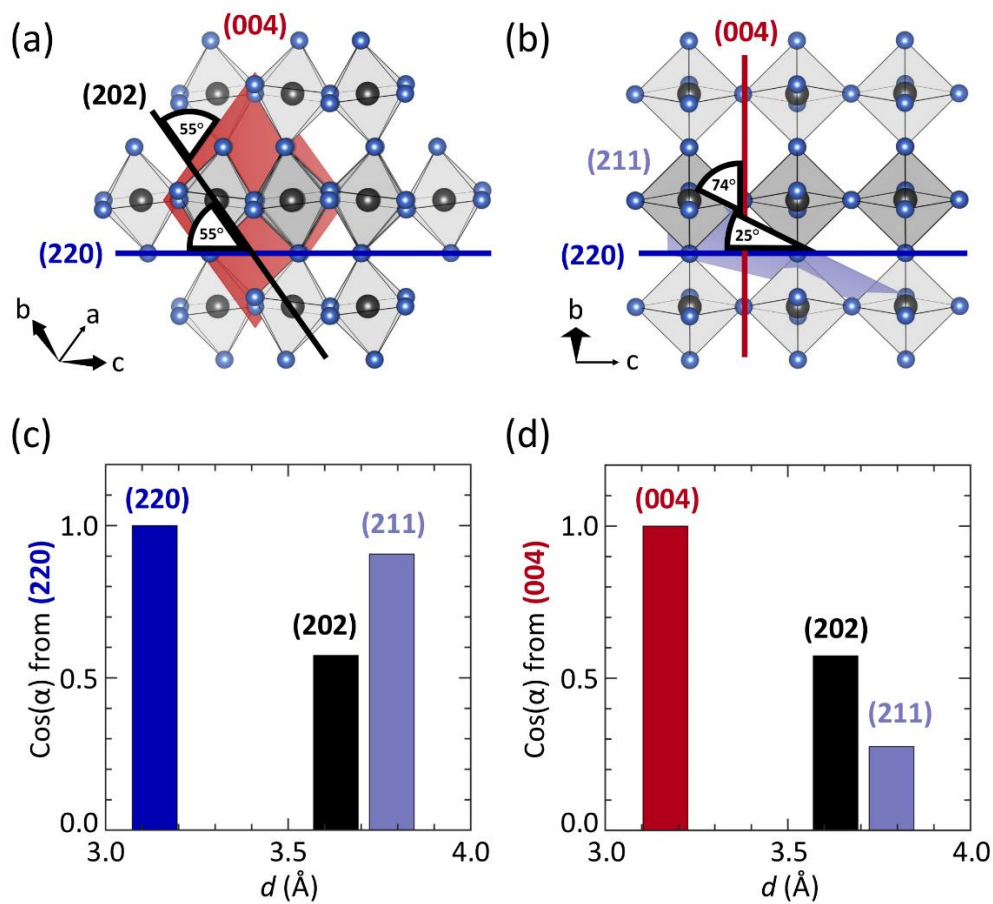

**Figure S3.** Crystal structures of MAPbI$_3$ in the tetragonal I4/mcm phase [2] showing (a) the angles made between the (202) plane and the (220) and (004) and (b) the angles made between the (211) plane and the (220) and (004). (c-d) Cosines of the angles in (a) and (b) respectively.

In order to properly assign GIWAXS peaks as (220) vs. (004), we tracked the intensity of the (211) peak with respect to its neighboring (202) peak. **Figures S3a-b** show the angles that the (202) (**Figure S3a**) and (211) (**Figure S3b**) planes make with the (220) and (004) planes. Conveniently, the (202) plane is at equal angle from the (220) and (004), making normalization of the (211) with respect to the (202) useful for analysis. The cosines of the above-mentioned angles are shown in **Figure S3c-d**. Because the (211) plane is nearly parallel to the (220), much stronger (211) intensity will be observed when the peak at $d \approx 3.1$ Å is (220) (**Figure S3c**). Correspondingly, because the (211) is nearly perpendicular to the (004), a weak (211) intensity with respect to the neighboring (202) peak likely indicates that the peak at $d \approx 3.1$ Å is (004) (**Figure S3d**). So, in nOP patterns, strong (211) correlates more with A domains and weak (211) correlates more with C domains. To further verify these assignments, we compared the nOP patterns with the nIP patterns, as a domain with (220) reflections out-of-plane should have corresponding (004) reflections in-plane (**Figure 1b**).

# Additional information regarding fraction *f*

To calculate *f*, the fitted peak intensities were corrected for structure factor, [58,59] and we used **Equation 1**:

$$f = 100 \times \frac{I_{Corr, peak\ of\ interest}}{I_{Corr, total}} \quad (1)$$

Where $I_{Corr}$ is the corrected peak intensity.

*f* represents the fraction of the film having (220) or (004) reflections, in the nOP *or* the nIP patterns. Thus, it is important to correct the scattering intensities of (220) and (004) relative to each other. The intensities $I_{hkl}$ of a peak (*hkl*) ((220) or (004)) were therefore corrected for structure factor $F_{hkl}$, using $I_{Corr} = (I_{hkl})(F_{hkl})^2$. [3] Conveniently, the (220) and (004) peaks were always within ≤ 0.1 Å of each other, and we calculated the *f*s for each *for the same nOP or nIP pattern*. Other correction factors [3] that might otherwise be necessary were therefore ignored, such as the Debye-Waller thermal correction or Lorentz/polarization factors. In addition, because each pattern probed specific domain orientations and not a powder sample, the multiplicity of all peaks should be 1. Thus, the corrected intensities were $I_{Corr} = (I_{hkl})(F_{hkl})^2$.

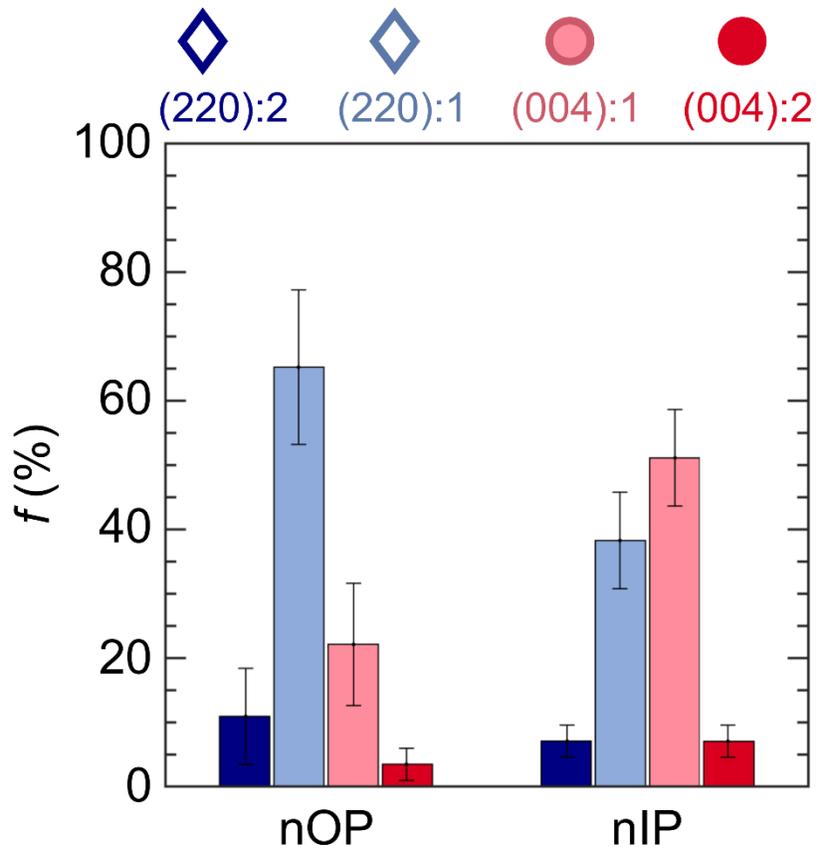

**Figure S4.** Reproduction of **Figure 3c** for 4 samples, where the values are the average *f*s and the error bars represent the standard deviation of these values. The results reproduce those shown in **Figure 3**, albeit with larger uncertainty values.

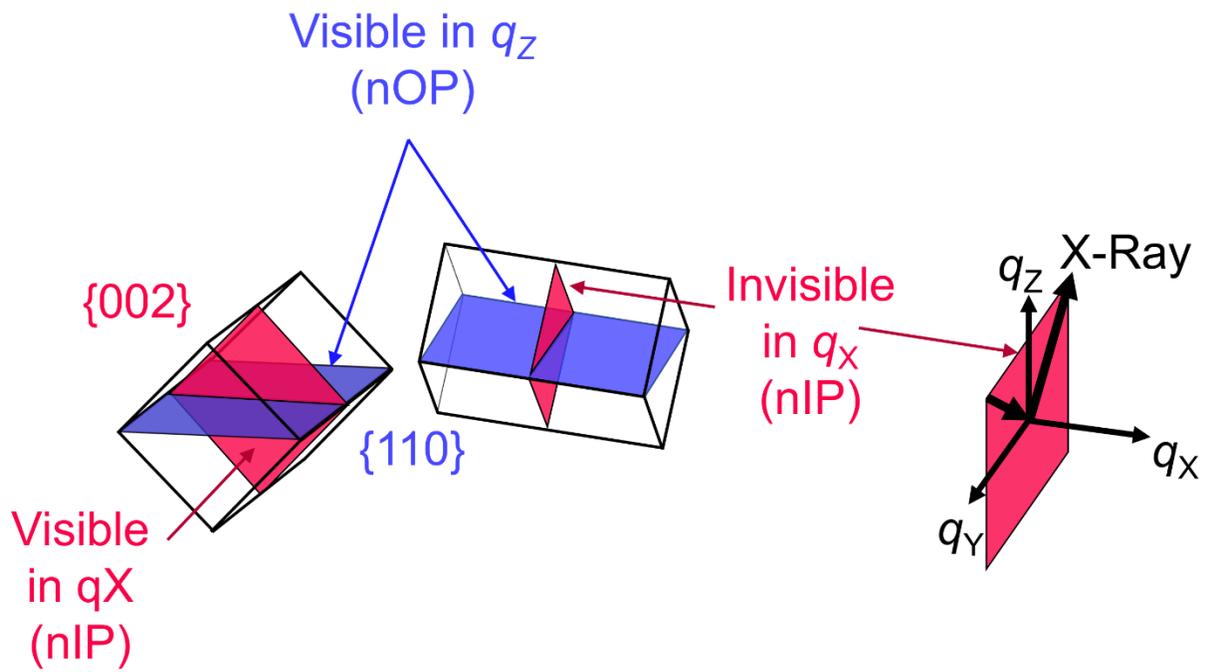

**Figure S5.** A-type domains with all planes visible nOP but with only some planes visible nIP, depending on domain rotation.

# Derivation of Equation for Minimum Domain Size $D_{min}$

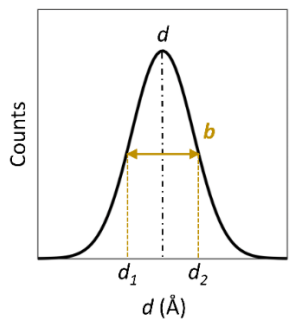

**Figure S6**. Sample peak in real-space, showing a peak of center $d$, with breadth $b$ spanning from $d_1$ to $d_2$.

Here, we extracted peak breadths to obtain domain size. Because we did this only for one peak, rather than over a series of peaks, the numbers reported in this study should be taken as the *minimum* domain size. We did not calculate size from a series of peaks because the (110) and (002) peaks significantly overlapped with each other and the (330) and (006) peaks significantly overlapped with Kapton peaks, so the (220) and (004) peaks were best for breadth calculations. We were also primarily concerned with *relative changes* to $D_{min}$ upon application of stress, not with the absolute values of the numbers themselves.

Prior work modified the Scherrer equation for GIWAXS to obtain the following: [4]

$$D_{hkl} = \frac{2\pi K}{\Delta q_{hkl}}$$

Where $D_{hkl}$ is the average domain size, $K$ is an *hkl*-dependent constant [5] and $\Delta q_{hkl}$ is the breadth of the Bragg reflection *hkl* in reciprocal space. However, because we analyzed all our data in real-space (*d*-spacings) and not in reciprocal space (*q*), we converted the above equation to real-space. Because $q = 2\pi/d$ and $\Delta q_{hkl} = q_1 - q_2$:

$$\Delta q_{hkl} = 2\pi \left(\frac{1}{d_1} - \frac{1}{d_2}\right) = 2\pi \left(\frac{d_2 - d_1}{d_1 d_2}\right) = 2\pi \left(\frac{b}{d_1 d_2}\right)$$

Where $b$ is the breadth (area divided by intensity)[4] in real space, $d$ is the center of the *hkl* peak, $d_1 = d - b/2$ and $d_2 = d + b/2$ (see **Figure S6).** Thus,

$$\Delta q_{hkl} = 2\pi \left( \frac{b}{(d - b/2)(d + b/2)} \right) = 2\pi b \left( \frac{1}{d^2 - b^2/4} \right)$$

$$D_{min} = \frac{K}{b} \left( d^2 - b^2/4 \right)$$

Where $K$ is the Scherrer constant ($\approx$ 1.0 for both (220) and (004))[5], $b$ is the breadth which in this case is the peak area divided by the peak intensity (uncorrected), and $d$ is the $d$-spacing. Because the same fitting methodology was consistently applied to all samples (Pseudo-Voigt with consistent Gaussian and Lorentzian contributions), we were able to extract trends in $D_{min}$ evolution, and thus draw conclusions about domain growth or shrinking. Conditions resulting in twin domain growth, shrinking and the implications for domain wall creation and annihilation are given in the discussion of **Figures 6**, **7**, **8** and **9**.

# Calculation of Applied Strain Magnitudes

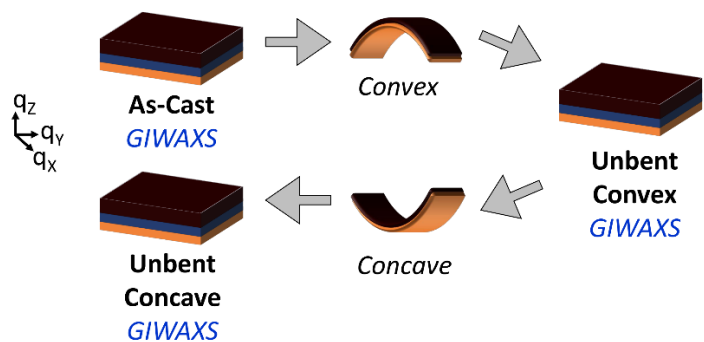

**Figure S7.** Bending experiment for convex and concave bending. GIWAXS patterns were taken after the applied strain was released.

Here we calculate the approximate magnitudes of the strains $\varepsilon$ applied to MAPbI$_3$ films during bending (**Figure S7**). To do so, we use the equation provided by Suo *et al.*:[6]

$$\varepsilon_{top,\ qx} = \left(\frac{t_f + t_S}{2R}\right) \frac{(1 + 2\eta + \chi\eta^2)}{(1 + \eta)(1 + \chi\eta)}$$

In this equation, the "top" designates the top region of the Kapton-PEDOT: PSS-MAPbI$_3$-PMMA stack; thus, the area closest to MAPbI$_3$. "$qx$" designates the strain in the $q_X$ direction; i.e. in-plane. $R$ is the radius around which the film is bent (so $2R$ is the diameter), $t_f$ is the film thickness, $t_S$ is the substrate thickness, $\eta = t_f/t_S$ and $\chi$ is the ratio of elastic moduli $Y$, where $\chi = Y_f/Y_S$.

The Young's modulus of MAPbI$_3$ has been found to vary between 10 and 20 GPa ($\approx$ 10, 12.8, 14, and 20 GPa);[7–10] so we take 14 GPa here as a mid-range value for approximation. The moduli of PEDOT: PSS is $\approx$ 2 GPa [11], that of PMMA is 3 GPa and that of Kapton is 2.5 GPa. Because the moduli of PEDOT: PSS and Kapton are not only very similar but also an order of magnitude lower than that of MAPbI$_3$, we can approximate the "substrate" as being composed of Kapton and PEDOT: PSS. In addition, because the modulus of PMMA is much lower than that of MAPbI$_3$ and because the PMMA layer is much thinner than the MAPbI$_3$ layer, we can approximate the "film" as being MAPbI$_3$ only. Thus, $\chi = Y_f/Y_S \approx Y_{MAPbI_3}/Y_{Kapton} \approx$ (14 GPa) / (2.5 GPa) $\approx$ 5.6.

Next, from **Figure S1**, we obtained the thickness of Kapton $t_{\text{Kapton}}$ (125 000 nm), of PEDOT: PSS ($\approx$ 200 nm), of MAPbI$_3$ ($\approx$ 400 nm) and of PMMA ($\approx$ 50 nm). Assuming again that the "substrate" is Kapton + PEDOT: PSS and that the "film" is MAPbI$_3$, $\eta = t_f/t_S \approx$ (400 nm)/(125200 nm) $\approx$ 0.00319. Because $\eta \sim 10^{-3}$, $\varepsilon_{top,\ qx}$ is practically independent of both $\eta$ and $\chi$.[6] We can therefore approximate the applied strain in the $q_X$ direction:

$$\varepsilon_X \approx \left(\frac{t_f + t_S}{2R}\right)$$

The $\varepsilon_X$ obtained are listed in **Table 1**. Then, Poisson's ratio was used to calculate $\varepsilon_Z$.

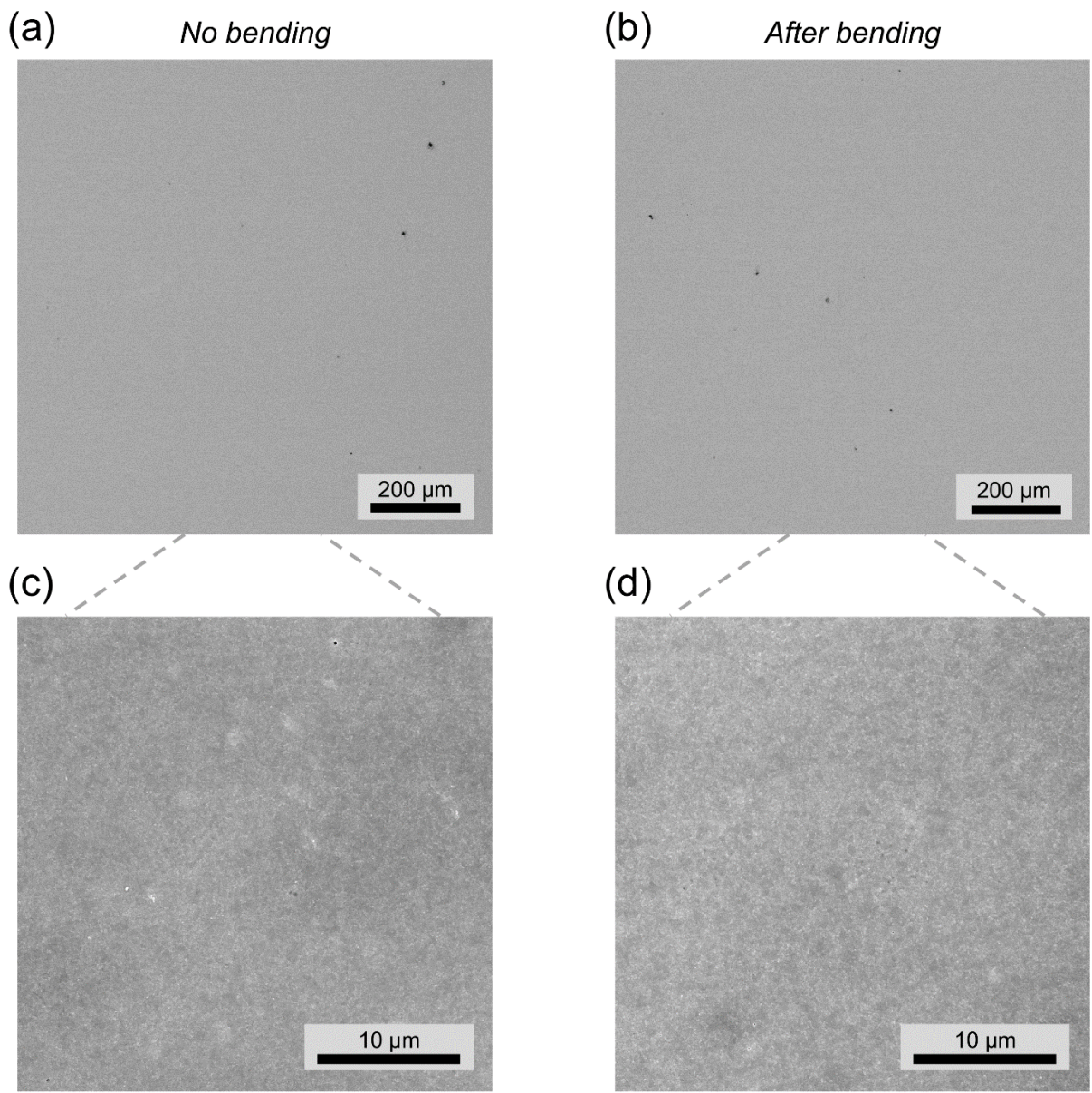

**Figure S8.** SEM micrographs of films with no bending ((a), (c)) and after bending 2 times ((b), (d)) *convexly* around a diameter of 4.1 mm, with low and high magnifications. No cracks were observed here. (19 keV)

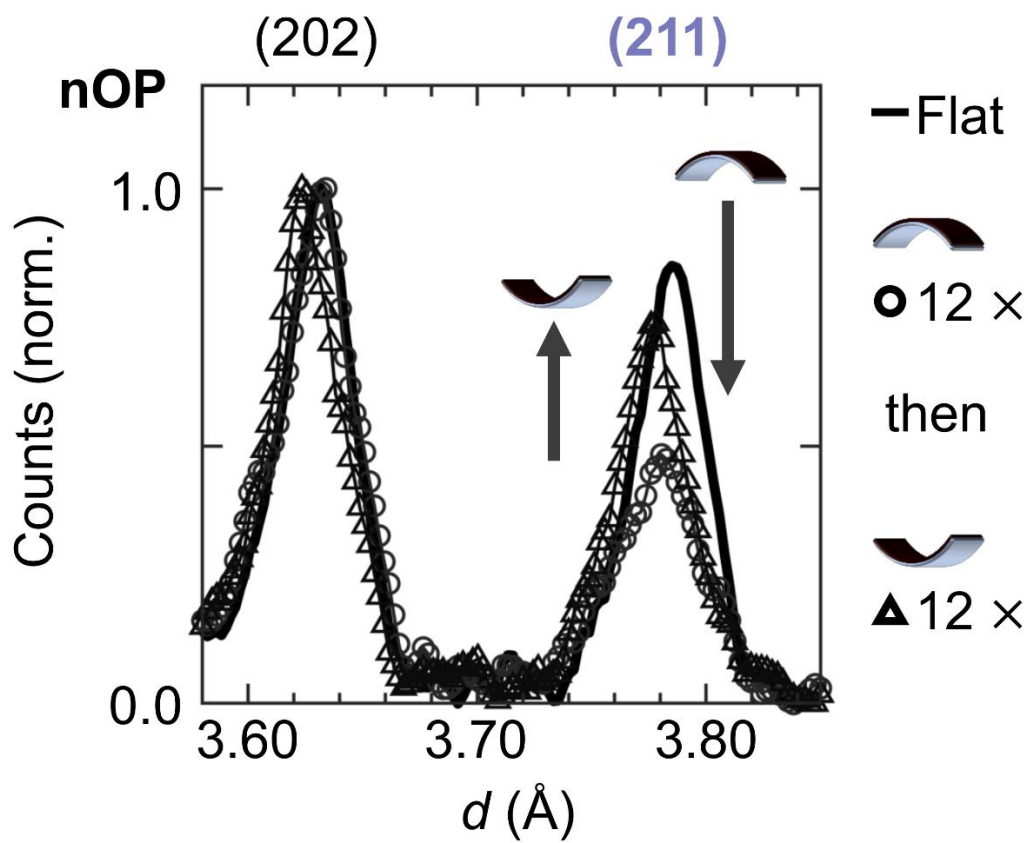

**Figure S9.** (211)-(202) region of the nOP patterns in **Figure 6** (repeated bending around 10 mm diameter), showing the expected drop in (211) intensity as C replaces A after *convex* bending, with subsequent recovery of the (211) (and of A) after *concave* bending.

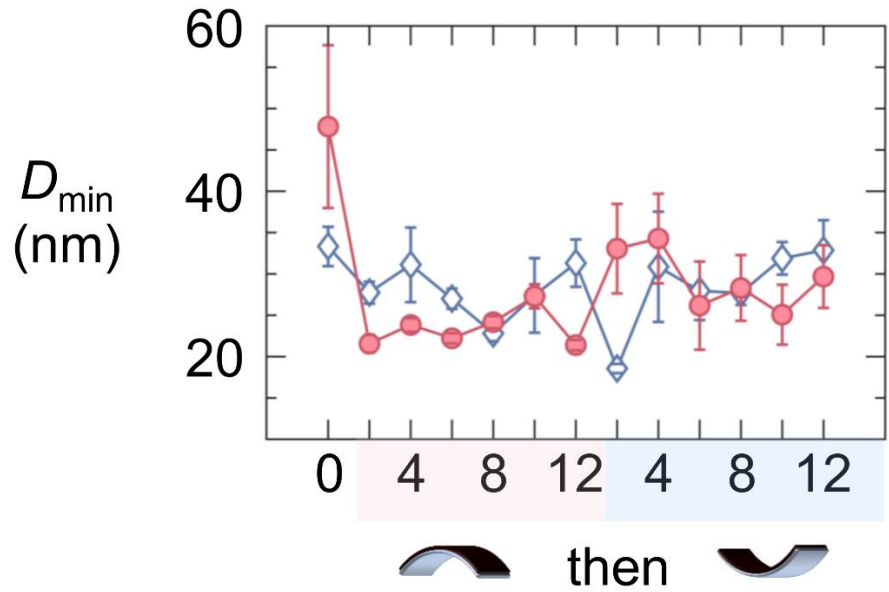

**Figure S10.** Changes in minimum domain size $D_{min}$ for the (220):1, and (004):1 peaks of the nOP patterns (see **Figure 6**) for the 10 mm bending experiment.

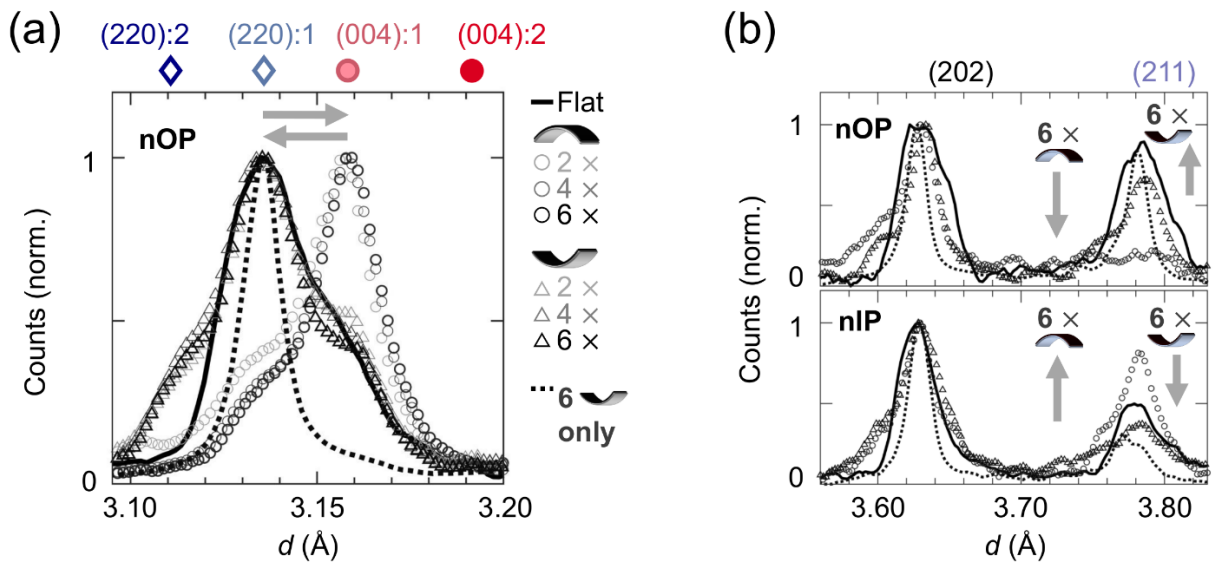

**Figure S11.** (a) GIWAXS nOP patterns in the (004)-(220) region for the successive *convex* and *concave* bending around a 4.1 mm diameter, as listed. (b) (221)-(202) region of select GIWAXS nOP and nIP linecuts.

*Convex* bending resulted in very large growth of the (004):1, with the main as-cast peak considerably decreased (**Figure S11a**). Correspondingly, the (211) strongly increased nIP and decreased nOP (**Figure S11b**). Subsequent *concave* bending resulted in an increase of the initial (220):1 peak, but with much (004):1 left over from the *convex* bending, and with a large and visible (220):2 shoulder. As expected, the (221) increased nOP and decreased nIP. Interestingly, when only *concave* bending was applied with no prior *convex* bending (called "*concave* bending only"), the (220):1 was practically the only peak observed, indicating mostly A:1. These striking changes are clear signs that much domain switching occurred during bending, with modifications made to the film being significantly retained when the applied strain was relieved.

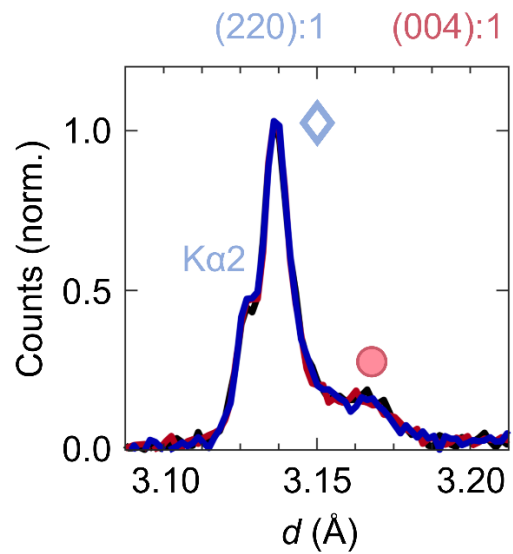

**Figure S12.** Three sequential powder XRD patterns (black, then red, then blue) of an as-cast MAPbI$_3$ film in the (220)-(004) region, showing a lack of change, and thus, stability under the beam.